\documentclass[twocolumn,superscriptaddress,longbibliography,aps,prb,showkeys]{revtex4-2}

\usepackage[normalem]{ulem}
\usepackage{graphicx}
\usepackage{bm}
\usepackage{color}
\usepackage{epstopdf}
\usepackage{amsmath}
\usepackage{amssymb}
\usepackage{lipsum}
\usepackage[ansinew]{inputenc}
\usepackage{multirow}
\usepackage{rotating}
\usepackage[urlcolor=blue,colorlinks=true,citecolor=blue,linkcolor=blue,pdfstartview={FitH},bookmarks=false]{hyperref}

\usepackage{xcolor}
\usepackage[OT4]{fontenc}
\begin{document}
\title{Tuning the topological character of half-Heusler systems: A comparative study on
Y$T$Bi ($T$ = Pd, Pt)}

\author{J. C. Souza}
\altaffiliation[Current address: ]{Department of Condensed Matter Physics, Weizmann Institute
of Science, Rehovot, Israel}
\affiliation{Instituto de F\'{\i}sica ``Gleb Wataghin'', Universidade Estadual de Campinas
(UNICAMP), 13083-859 Campinas, SP, Brazil}
\affiliation{Max Planck Institute for Chemical Physics of Solids, D-01187 Dresden, Germany}

\author{M. V. Ale Crivillero}
\affiliation{Max Planck Institute for Chemical Physics of Solids, D-01187 Dresden, Germany}

\author{H. Dawczak-D\k{e}bicki}
\affiliation{Max Planck Institute for Chemical Physics of Solids, D-01187 Dresden, Germany}

\author{Andrzej Ptok}
\affiliation{Institute of Nuclear Physics, Polish Academy of Sciences, W. E. Radzikowskiego
152, PL-31342 Krak\'{o}w, Poland}

\author{P. G. Pagliuso}
\affiliation{Instituto de F\'{\i}sica ``Gleb Wataghin'', Universidade Estadual de Campinas
(UNICAMP), 13083-859 Campinas, SP, Brazil}
\affiliation{Los Alamos National Laboratory, Los Alamos, New Mexico 87545, USA}

\author{S. Wirth}
\email[e-mail: ]{steffen.wirth@cpfs.mpg.de}
\affiliation{Max Planck Institute for Chemical Physics of Solids, D-01187 Dresden, Germany}

\date{\today}

\begin{abstract}
Half-Heusler systems host a plethora of different ground states, especially with non-trivial
topology. However, there is still a lack of spectroscopic insight into the corresponding
band inversion in this family. In this work, we locally explore the half-Heuslers Y$T$Bi
($T =$ Pt and Pd) by means of scanning tunneling microscopy/spectroscopy. From our analysis
of the (120) surface plane, we infer that the increase of the spin--orbit coupling upon
going from Pd to Pt is the main player in tuning the surface states from trivial to
topologically non-trivial. Our measurements unveil a ($2 \times 1$) reconstruction of the
(120) surface of both systems. Using density functional theory calculations, we show that
the observed different behavior of the local density of states near the Fermi level in these
two materials is directly related to the presence of metallic surface states. Our work sheds
new light on a well known tunable family of materials and opens new routes to
explore the presence of topological states of matter in half-Heusler systems and its
microscopic observation.
\end{abstract}

\maketitle

\section{Introduction}
The seminal works on the quantum (spin) Hall effect \cite{klitzing1980new,
bernevig2006quantum,konig2007quantum} were crucial to definitely incorporate topology 
into the analysis of electronic band structure of solids \cite{manna2018heusler,
armitage2018weyl,hasan2010colloquium}. The net result was the prediction and observation
of a plethora of quantum topological states of matter, such as topological insulators 
(TIs) \cite{hasan2010colloquium}, Dirac and Weyl semimetals \cite{armitage2018weyl,
lv2021experimental} and even more exotic excitations \cite{wang2016hourglass,
xu2012hedgehog}. Due to its unique physical properties, in which surface states often 
play a decisive role \cite{hasan2010colloquium,armitage2018weyl}, the application of 
such systems can reach from spintronics to quantum computing \cite{tokura2017emergent}.

Despite such potential applications, two key ingredients have been limiting factors for a
broader use of TIs \cite{tokura2017emergent,singh2022topology}. The first one is that many
materials have their Fermi energy $E_{\rm F}$ located in one of the bands derived from the
bulk. In other words, experimentally the bulk is not fully insulating, as, e.g. typically
observed in layered chalcogenides \cite{singh2016role,eguchi2015precise,eremeev2012atom,
singh2022topology}. Secondly, the Dirac point is often located sizeably away from $E_{\rm F}$,
preventing these materials from potential usage in, e.g., transport applications. A solution
for both problems may reside in correlated systems, where the many body interactions pin the
Dirac point close to $E_{\rm F}$, within the bulk gap \cite{jiao2018,pirie2020}. However,
their correlated phases normally appear only at low temperatures, which implies that those
topological phases may not be suitable for applications \cite{guo2018evidence,dzsaber2021giant,
aishwarya2022spin,wirth2021stm,paschen2021quantum}. As such, it is imperative to find
appropriate materials with an insulating-like bulk and Dirac points near $E_{\rm F}$, whose
properties can also be tuned to specific requirements and even show a good match to important
semiconducting substrates \cite{palmstrom2003epitaxial}.

\begin{figure}[t]
\includegraphics[width=0.75\columnwidth]{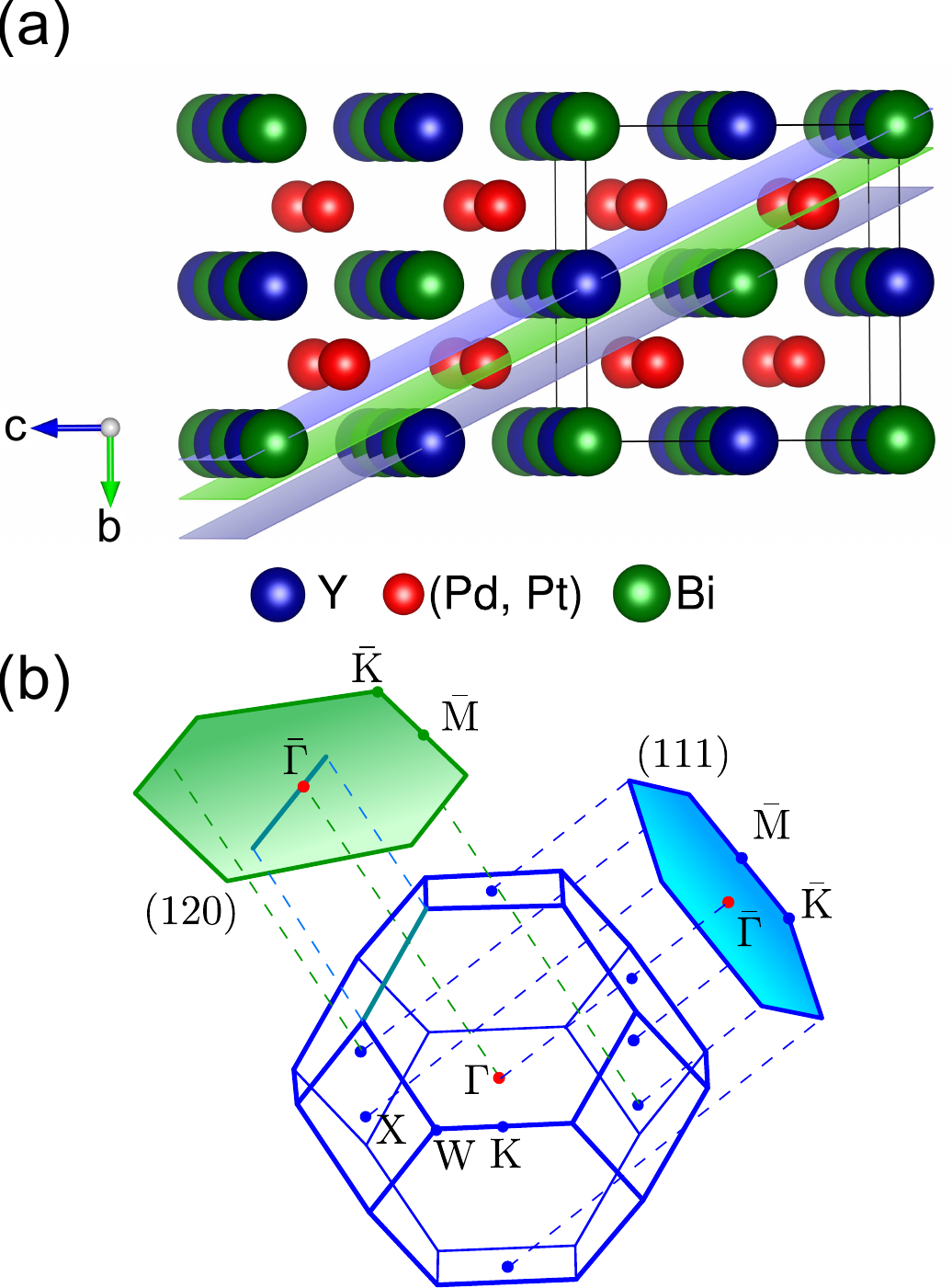}
\caption{(a) Crystal structure of the half-Heusler systems. The blue and green planes
indicate the (120) planes of YBi and Pd/Pt termination, respectively. The black lines
outline the unit cell. (b) Brillouin zone and its projection for the (111) and (120)
surface planes.} \label{Fig1}
\end{figure}

One of the most versatile systems that host numerous topological states of matter is the
half-Heusler compounds \cite{manna2018heusler,graf2011simple}. This family, with a simple
MgAgAs-type cubic structure (space group $F\overline{4}3m$), that can be seen as a ZnS-type
structure with filled octahedral lattice sites [Fig. \ref{Fig1} (a)] \cite{graf2011simple},
has been extensively explored due to the fact that its semiconducting, magnetic, thermoelectric
and strongly correlated properties can often be tailored \cite{zeier2016engineering,
mun2013magnetic,manna2018heusler,hundley1997electronic,fisk1991massive,eriksson1992electronic}.
Earlier theoretical calculations have suggested that the mechanism behind the appearance of
topological features in this family depends on the band inversion, which is very similar to
the one observed in the prototypical CdTe and HgTe systems \cite{feng.xiao.10}. In both
compounds, at the $\Gamma$ point near $E_{\rm F}$, the energy bands are split into
$\Gamma_{6}$, $\Gamma_{7}$ (both twofold degenerate) and $\Gamma_{8}$ (fourfold degenerate)
states. This splitting originates from the zinc blende crystal symmetry and strong spin-orbit
coupling \cite{feng.xiao.10,al2010topological,chadov2010tunable,lin2010half}. In the context
of band topology, CdTe possesses a normal band order (the $s$-like $\Gamma_{6}$ state sits
above the $p$-like $\Gamma_{8}$ state) while in HgTe band inversion occurs such that
$\Gamma_{6}$ resides below $\Gamma_{8}$ \cite{zhu2012band}. Both situations are reproduced in
Y$T$Bi, where for $T$ = Pd the normal (trivial) state is realized, while for $T$ = Pt a band
inversion occurs and non-trivial states emerge (see Fig.\ \ref{bulkband} in the Appendix) \cite{feng.xiao.10,al2010topological,chadov2010tunable,lin2010half}.

In fact, previous angle resolved photoemission spectroscopy (ARPES) measurements on YPtBi
have shown the presence of unusual topological surface states \cite{liu2016observation,
hosen2020observation}. Here, the topological surface states are observed throughout the
Brillouin zone. This situation is different from the one typically observed, e.g., in
the chalcogenides TIs, where the topological surface states emerge as a Dirac cone
\cite{king.hatch.11,liu2016observation,zhu2012band,ruan2016symmetry}. Moreover, topological
features were also observed in form of Weyl fermions in related half-Heusler semimetals
$R$PtBi ($R$ = Nd, Gd, Yb) \cite{hirschberger2016chiral,guo2018evidence,shekhar2018anomalous}.
In particular, topological features can affect a possible superconducting state, even
resulting in triplet superconductivity for some members of the $R$$T$Bi family ($R$ = rare
earth, $T$ = Pd or Pt) \cite{brydon2016pairing,butch2011superconductivity,gofryk2011magnetic,
pavlosiuk2016superconductivity,nakajima2015topological,kim2021campbell,kim2018beyond}.

In an even more fundamental aspect, comparing YPtBi and YPdBi can be an excellent platform
to experimentally tune the topological properties through the spin--orbit coupling. Both
systems possess very similar lattice parameters [6.652(1) {\AA} and 6.639(1) {\AA} for YPtBi
\cite{butch2011superconductivity} and YPdBi \cite{gofryk2011magnetic}, respectively], which
makes the spin--orbit coupling the key parameter to distinguish between trivial (YPdBi) and
non-trivial (YPtBi) topological states \cite{al2010topological,chadov2010tunable,lin2010half,
liu2016observation,hosen2020observation}. These compounds are particularly attractive due
to the possibility of obtaining high quality thin films, increasing their potential
applicability \cite{kim2022molecular}. Previous nuclear magnetic resonance (NMR)
\cite{nowak2014nmr,zhang2016nmr} and electron spin resonance (ESR) experiments
\cite{lesseux2016} pointed toward a strong impact of spin--orbit coupling on the detailed
band structure of Y$T$Bi. A direct experimental visualization of the surface states
resulting from band inversion in half-Heuslers has not been demonstrated, yet.

Additionally, although this family of materials supports so many different physical
properties, which are often related to surface states, little is known about the surface
properties of half-Heuslers. This is, at least in part, certainly related to the fact
that half-Heusler compounds with cubic structure are notoriously difficult to cleave,
rendering an {\it in situ} preparation of atomically flat surfaces from bulk samples a
challenge. In consequence, reports employing scanning tunneling microscopy/spectroscopy
(STM/STS) are scarce and focused on disordered surfaces in single crystals
\cite{baek2015creating} or on the study of surface reconstructions in thin films
\cite{kawasaki2018simple}. In this work, we report on atomically flat surfaces
investigated by STM/STS, combined with first-principles density functional theory (DFT)
slab calculations, to explore the local properties of the half-Heuslers YPdBi and
YPtBi. We cleaved our samples {\it in situ}, most likely along the (120) planes exposing a
($2\times1$) reconstructed YBi-terminated surface. From our STS data we infer a \emph{finite}
local density of states (local DOS or LDOS) $\eta$($E$) at $E_{\rm F}$ for YPtBi, while the
trivial YPdBi compound exhibits a well defined gap around $E_{\rm F}$. We argue that the
difference in the LDOS is likely due to the formation of metallic surface states in
YPtBi, a finding corroborated by our slab calculations. Our work establishes the possibility
of using STM as a local probe to investigate half-Heusler systems and suggests that a
tuning of the LDOS can be achieved through the increase of the spin--orbit coupling
upon going from Pd to Pt.

\section{Methods}
\label{sec:Methods}
Single crystalline samples of Y(Pd,Pt)Bi were synthesized by the Bi self flux growth
technique with starting elements Y (99.99\%):(Pd,Pt) (99.99\%):Bi (99.999+\%) in the
proportion of 1:1:10 \cite{souza2019crystalline}. While YPtBi samples naturally
expose (001), (110) and (111) planes in a pyramid-like shape, YPdBi samples only
expose (001) planes (all samples had a cube-like shape). The investigated samples had
an approximate size of $1\times1\times$1 mm$^{3}$.

STM/STS measurements were performed in a ultrahigh vacuum system at pressures
$p \leq 2.5 \times 10^{-9}$~Pa and at temperatures $T = 4.6$~K. A total of seven
(four YPdBi and three YPtBi) samples were cleaved {\it in situ} at temperatures $T
\approx 20$~K. The tunneling current $I$ was measured using electrochemical prepared
tungsten tips and a bias voltage $V_{b}$ was applied to the samples. The topographies
were obtained in a constant current mode with a predefined current set point $I_{sp}$.
Most topographies were obtained in a dual-bias mode, i.e., forward and backward scans
along the fast scan direction were obtained with a $V_{b}$ of the same magnitude, but
with opposite signs. We did not see any differences in dual-bias mode (i.e.\ for the
different values of $V_b$) when scanning the samples along the (120) planes. The
d$I$/d$V$-spectra were acquired by a lock-in technique applying a modulation voltage
of typically $V_{mod} = 0.3$~mV at $117$~Hz.
\begin{figure*}[t]
\includegraphics[width=\textwidth]{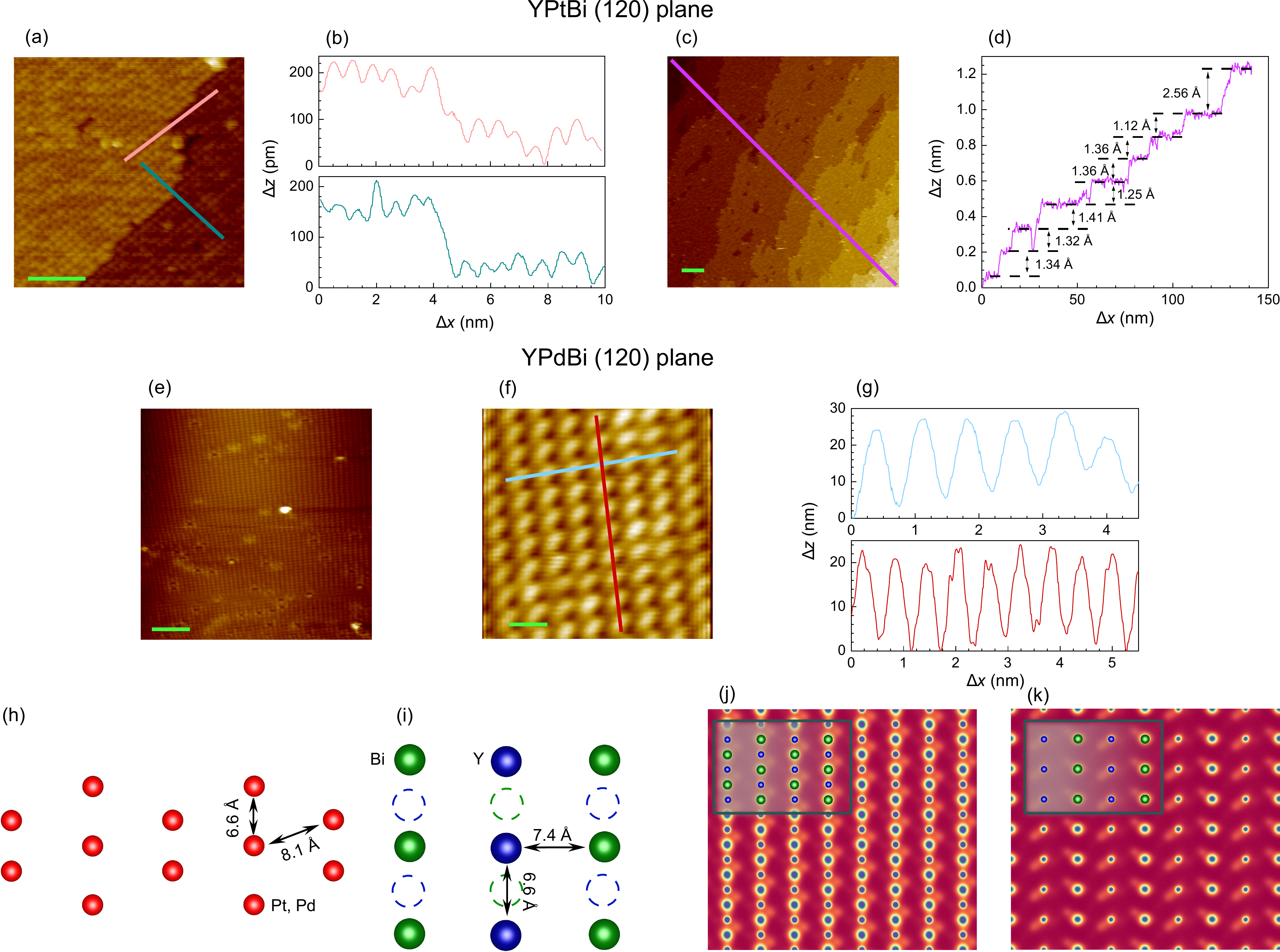}
\caption{(a) $20 \times 20$~nm$^{2}$ topography along the (120) plane for YPtBi (bias
voltage $V_{b} = - 300$~mV, $I_{sp} = 0.7$~nA, scale bar of $5$~nm). The height scans
along two (almost) perpendicular directions are shown in (b) by curves of corresponding
color. The step edge height is, approximately, $115$~pm. (c) $100 \times 100$~nm$^{2}$
field of view (scale bar of 10~nm). The height scan along the violet line over several
step edges is presented in (d). (e) $30 \times 30$~nm$^{2}$ STM topography along the
(120) plane for YPdBi ($V_{b} = - 200$~mV, $I_{sp} = 0.6$~nA, scale bar of $5$~nm). (f)
High resolution $6 \times 6$~nm$^{2}$ field of view (scale bar of $1$~nm). (f) Two
perpendicular height scans from (f) are presented by corresponding colors. View onto the
topmost layer of (h) Pd/Pt- and (i) YBi-terminated surfaces for the (120) plane. In (i),
the proposed ($2\times1$) surface reconstruction is indicated where hollow circles mark
empty atom positions. Theoretically predicted STM topography for (120) surfaces: (j)
unreconstructed and (k) ($2\times1$) reconstructed YBi-terminated surface. The
topography was simulated for a tip $\sim$1~\AA\ above the surface of area $5.3 \times
5.9$~nm$^{2}$. The blue (red) color corresponds to high (low) charge density.}
\label{Fig2}  \end{figure*}

The first-principles density functional theory (DFT) calculations were performed using
the projector augmented-wave (PAW) potentials \cite{blochl.94} implemented in the
Vienna Ab initio Simulation Package ({\sc vasp}) code \cite{kresse.hafner.94,
kresse.furthmuller.96,kresse.joubert.99}. The calculations containing the spin--orbit
coupling (SOC) were performed with the generalized gradient approximation (GGA) under
the modified Becke-Johnson (mBJ) exchange potential~\cite{becke.johnson.06,
tran.blaha.09,camargo.baquero.12}. The energy cutoff for the plane-wave expansion is
set to $350$~eV. The density of states was calculated using $12 \times 12 \times 12$
{\bf k}--point $\Gamma$--centered grids in the Monkhorst--Pack scheme
\cite{monkhorst.pack.76}. The lattice constants were assumed to be equal to the
experimental values, i.e. $\approx$ $6.64$~\AA\ for both compounds
\cite{haase.schmidt.02}. The band structures from the DFT calculations were used to
find tight binding models by {\sc Wannier90}~\cite{mostofi.yates.14,pizzi.vitale.20},
which allowed us to calculate surface state spectra by {\sc WannierTools}
\cite{wy.zhang.18}. The theoretical simulation of STM topographies for (120)
YBi-terminated surfaces (without and with reconstruction) were computed using the
Tersoff--Hamann approach \cite{tersoff.hamann.85}. Due to technical limitations of
the mBJ potential for slab-type calculations, these specific DFT calculations were
performed using a GGA with Perdew--Burke--Ernzerhof (PBE) parametrization
\cite{perdew.burke.96}. More details on the calculations are provided in the
Appendix \ref{appendD}, {\it Details of Bulk and surface band structure calculations}.

\section{Results and Discussion}
\subsection{Topography of YPtBi single crystals}
\label{sec:topoPt}
{\it In situ} preparation of clean surfaces (in case of bulk single crystals typically
by cleaving) is of utmost importance for STM/STS studies but often exceedingly difficult for
materials of cubic crystal structure \cite{wirth2021stm}. Our single crystals of half-Heusler
compounds YPdBi and YPtBi naturally expose (001) crystallographic planes, while the (111)
plane was only found for YPtBi. Figure \ref{Fig1}(b) shows the surface projection of the
Brillouin zone along the latter direction. In principle, the pyramid-like shape of our
YPtBi samples may open the possibility of exploring surfaces along the (001) and (111)
planes. However, the YPdBi crystals had a more cube-like shape, suggesting a preferred
cleave along the (001) plane. We emphasize that for a reasonable comparison between results
obtained on both compounds it is vital to investigate \emph{identical} crystallographic
planes. Therefore, we focus on samples mounted along the (001) direction in the following
(further details of measurements for cleaving YPtBi along the (111) plane are provided in
Appendix \ref{appendA}, see Figs.\ \ref{S111}, \ref{topo111} and \ref{step111}).

Atomically flat surfaces on cleaved half-Heuslers are extremely scarce and required extensive
search. One example is shown in Fig.\ \ref{Fig2}(a) for YPtBi where, in principle, the
cleaving was expected to occur along the (001) plane. Before comparing the results from two
different compounds, it is crucial to identify which planes and terminations are obtained
since surface states depend decisively on those two parameters \cite{liu2016observation,
hosen2020observation}. Notably, all the obtained flat surfaces show a $\sim 30^{\circ}$
tilt angle with respect to the sample mounting plane (001) in this cleaving configuration.
This is a hint that the obtained surfaces are \emph{not} (001) planes. As will be argued
below, the exposed planes are likely YBi-terminated (120) planes instead, which are
highlighted in blue in Fig.\ \ref{Fig1}(a).

Three crucial pieces of information are helpful to identify the cleaving plane: (i) As
mentioned, the surfaces are tilted by $\sim 30^{\circ}$ with respect to the sample mounting
plane. This renders the (120) plane a likely surface as it is expected 26.6$^{\circ}$ away
from the (001) plane. (ii) The distance between corrugations and the (iii) the height of
the step edges can be analyzed. Fig. \ref{Fig2}(a) exemplifies the latter two in the same
field of view. The apparent height profile $\Delta z$ as a function of the lateral
distance $\Delta x$ (height scans) shown in Fig. \ref{Fig2}(b) reveals a step edge height
of $\sim 115$~pm and a distance between corrugations $d_{exp}^{(120)Pt} = 0.66(3)$ and
$0.68(3)$~nm for the pink and turquoise directions, respectively. Those distances are in
agreement with the ones extracted from the Fourier transform of larger areas, from which
we obtain $d_{exp}^{(120)PtFT} = 0.67(3)$ and $0.70(3)$~nm (see Fig.\ \ref{FT120} in the
Appendix).

In order to gain better statistics on the average step edge height a $100 \times 100$
nm$^{2}$ area obtained on YPtBi is investigated, Fig.\ \ref{Fig2}(c). We can clearly see
the occurrence of several step edges and a lack of adatoms. We note that the latter
observation distinguishes the here observed surfaces over measurements on the (111)
plane (compare Fig.\ \ref{S111} in the Appendix). The average step edge height between
different exposed surfaces is $d^{(120)}_{exp} \approx 135$~pm [Fig.\ \ref{Fig2}(d)]
indicating that either exclusively YBi-terminated or Pt-terminated are observed. Note
that the theoretically expected distance between these planes is 148~pm, see Fig.\
\ref{Fig1}(a). A Pt-terminated surface can be ruled out since a distorted hexagonal
lattice with distances between Pt atoms of 0.66 and 0.81 nm would be expected, Fig.\
\ref{Fig2}(h), which is in clear contrast to the observation of Fig.\ \ref{Fig2}(a). On
the other hand, for a YBi-terminated surface a rectangular lattice with distances of
0.33 and 0.74 nm between atoms is expected. Therefore, we propose a ($2 \times 1$)
reconstructed YBi surface where half of the atoms are missing, see Fig.\ \ref{Fig2}(i).
This scenario is consistent with the observed distances between corrugations and STM
simulations obtained through slab DFT calculations.

Reconstructed surfaces, including the $(2 \times 1)$ type, are commonly observed
on both, bulk samples \cite{kawasaki2018simple} and thin films \cite{bach2003,
Kawasaki-APL-2014,Patel-APL-2014,Patel-CrysGrowth-2016} of half-Heusler compounds.
The driving forces behind these reconstructions were argued to be charge neutrality
and a minimization of the number of dangling bonds \cite{kawasaki2018simple,
liu2016observation,hosen2020observation}. However, the (120) surface plane has not
been investigated so far. In order to get further insight, we conducted
first-principles slab calculations for this particular surface termination.
Specifically, the total energies for slabs without and with $(2 \times 1)$
reconstruction were calculated. The reconstructed slabs contained 84 sites, i.e.\ 28
atoms of each species (for further details see Appendix \ref{appendD}). To allow
comparison to the non-reconstructed surface, two Pd/Pt atoms were removed from one
surface, but added as free atoms to the total energies. The calculations clearly
favor a reconstructed surface by about 4.1 eV in case of YPdBi, and about 8.3 eV for
YPtBi.

From the valence situation in the half-Heusler compounds, one may expect an
YBi-terminated surface to be charge-neutral. In line with the statement above
\cite{kawasaki2018simple} one may then speculate about a minimum number of dangling
bonds for the reconstructed surface and hence, a limited impact of dangling bonds
on the surface properties.
\begin{figure*}[!ht]
\includegraphics[width=0.92\textwidth]{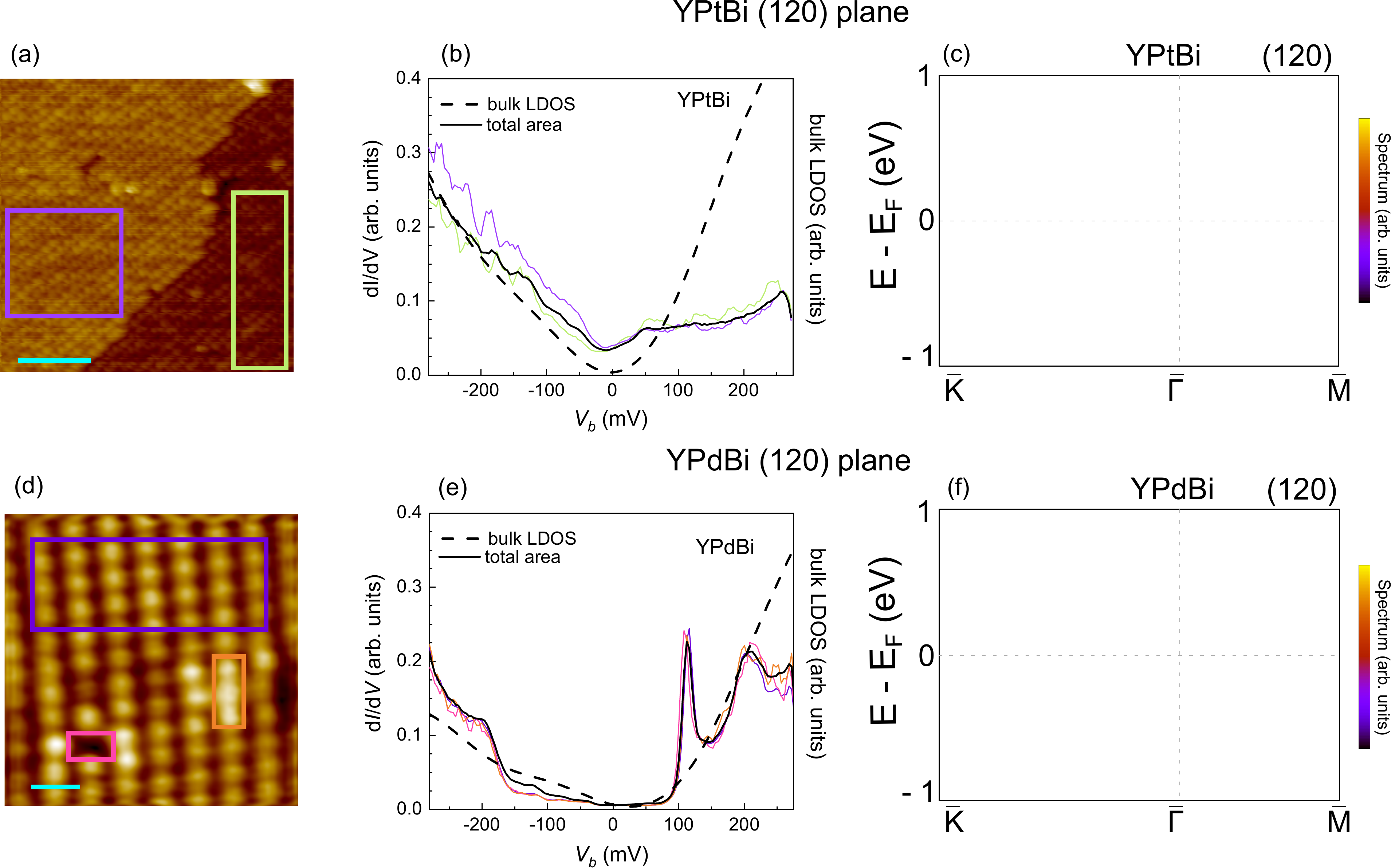}
\caption{(a) Topography of YPtBi as in Fig.\ \ref{Fig2}(a) with areas marked within which
d$I$/d$V$-spectra were averaged (in addition to the total area). (b) d$I$/d$V$-spectra
within the purple and green rectangles as well as for the total area (black) presented in
(a). Also, the calculated bulk DOS (dashed line) is included. The latter is normalized at
negative bias to the experimental value to improve the visualization. (c) Slab calculated
electronic band structure for a YBi-terminated (120) plane of YPtBi. (d) $6 \times
6$~nm$^{2}$ topography for YPdBi ($V_{b} = -200$~mV, $I_{sp}$ = 0.6 nA, scale bar of
$1$~nm). (e) d$I$/d$V$-spectra averaged over the magenta, orange, purple and the whole
area shown in (d). Again, the calculated bulk DOS for YPdBi is included for comparison
(dashed line). (f) Slab calculated electronic band structure for the YBi-terminated (120)
plane of YPdBi.}  \label{Fig3}
\end{figure*}

\subsection{Topography of YPdBi}
\label{sec:topoPd}
Naturally, also on YPdBi atomically flat surfaces needed to be extensively searched for.
Areas of $30 \times 30$~nm$^{2}$ could be identified, as exhibited in Fig. \ref{Fig2}(e).
However, we were not able to find any step edges in all of our investigated YPdBi cleaves.
Importantly, within these areas we observed the same rectangular pattern as for YPtBi.
This pattern is confirmed by high resolution topographies, as presented in Fig.
\ref{Fig2}(f), where the obtained distances between corrugations are $d_{exp}^{(120)Pd} =
0.61(3)$ and $0.72(3)$~nm [Fig. \ref{Fig2}(g)]. These values are consistent with results
from Fourier analyses obtained on bigger areas (see Fig.\ \ref{FT120} in the Appendix)
as well as with the YPtBi results.

Apart from the missing step edges, all of the investigated, atomically flat areas on
YPdBi appeared to be consistent with the plane orientation and termination as observed for
YPtBi cleaves. In particular, the experimentally obtained surfaces are again tilted by
$\sim 30^{\circ}$ with respect to the sample mounting plane (001). Hence, our observation
point again toward $(2\times 1)$ reconstructed surfaces along the (120) plane.

Figures \ref{Fig2}(j) and (k) represent STM simulations obtained through slab DFT
calculations for YBi-terminated surfaces without and with $(2\times 1)$ reconstruction,
respectively (for details, see Appendix \ref{appendD3}). These results indicate that
without a surface reconstruction we would likely observe stripes along the (100)
crystallographic direction [Fig.\ \ref{Fig2}(j)]. Such stripes are absent on ($2 \times
1$) reconstructed surfaces [Fig.\ \ref{Fig2}(k)], in line with our observations.

The simulations also help explaining the subtle differences in the topographies upon going
from Pd to Pt samples. For a (120) termination, the Pd/Pt atoms reside only 74 pm below
the topmost YBi layer and hence, the Pd/Pt atoms may also contribute to the topography, as
suggested by the yellow contributions in Figs.\ \ref{Fig2}(j) and \ref{Fig2}(k). The radial
extent of the 4$d$ orbitals is smaller than the 5$d$ ones \cite{tomczak2020isoelectronic,
ajeesh2022ground}. Therefore, the second-to-topmost layer may have slightly different
contributions to the topography depending on whether it is Pd or Pt.

\subsection{Bulk properties of Y$T$Bi ($T =$ Pd, Pt)}
\label{sec:bulk}
Before discussing our spectroscopic results of the \emph{surface} properties of YPdBi and
YPtBi, we address possible differences in the \emph{bulk} DOS near $E_{\rm F}$ for these
two materials as this may easily influence the spectral weight measured at the surface.

As already mentioned, half-Heusler systems have a three dimensional character and, as such,
bulk states might be relevant \cite{marques2021tomographic,rhodes2022nature}. In fact,
previous specific heat studies have obtained very similar Sommerfeld coefficients $\gamma$
for both systems. While for YPdBi $\gamma = 0.3(1)$ mJ$\,$mol$^{-1}$K$^{-2}$ was reported
\cite{souza2019crystalline}, results for YPtBi ranged from $\sim$0.1 to $0.4$
mJ$\,$mol$^{-1}$K$^{-2}$ \cite{pagliuso1999crystal,pavlosiuk2016superconductivity}.
Assuming a free conduction electron gas model with $\gamma = (2/3) \pi k_{B}^{2}
\eta_s(E_{\rm F})$, where $k_{B}$ is the Bohr magneton and $\eta_s(E_{\rm F})$ denotes
the spin-resolved DOS at $E_{\rm F}$, one obtains $\eta_{s}(E_{\rm F})^{\rm YPdBi} =
0.06(4)$~eV$^{-1}$\,mol$^{-1}$\,spin$^{-1}$ for YPdBi, and $\eta_{s}(E_{\rm
F})^{\rm YPtBi} = 0.04(4)$~eV$^{-1}$\,mol$^{-1}$\,spin$^{-1}$ for YPtBi. Such a negligible
electronic contribution to the specific heat is consistent with previous transport
measurements for both systems, which reported a semiconductor/semimetal-like behavior
\cite{butch2011superconductivity,pagliuso1999crystal,souza2019crystalline,wang2013large,
pavlosiuk2016superconductivity,gofryk2011magnetic,nakajima2015topological,souza2018diffusive}.

It is worth noting that Pd/Pt and Bi-based compounds are known for hosting impurity phases,
such as Bi and/or Pd/Pt-Bi binary phases \cite{ajeesh2022ground}. Such impurity phases may
affect the macroscopic properties, especially transport measurements. Consequently, it is
highly desirable to have an experimental confirmation of the insulating bulk nature from a
microscopic technique. Indeed, previous electron spin resonance measurements for rare-earth
substituted YPdBi and YPtBi clearly indicate an insulating bulk behavior. This establishes
the presence of a small gap in the bulk DOS at $E_{\rm F}$ of both systems
\cite{pagliuso1999crystal,souza2019crystalline}, in agreement with our DFT results discussed
below.

\subsection{Spectroscopic results on Y$T$Bi ($T =$ Pd, Pt)}
\label{sec:spec}
Having identified identical surface terminations and established negligible bulk
contributions to the DOS near $E_{\rm F}$ for both materials YPdBi and YPtBi, we can now
compare their surface electronic properties. In Fig.\ \ref{Fig3}(b) and (e) the STS
results, i.e.\ d$I$/d$V$-spectra, are presented. We note that, within simplifying
assumptions, d$I$/d$V \propto \eta$($E$). The topographic areas over which the spectroscopy
curves were averaged are shown in Figs.\ \ref{Fig3}(a) and (d), respectively, with the
black curves in (b) and (e) obtained within the total areas of (a) and (d). Clearly, there
are only minor differences between spectra obtained in different areas of a given compound.
In particular, for YPdBi also spectra obtained at different defects are included, see
orange and pink rectangles/curves in Fig.\ \ref{Fig3}(d) and (e), which do not
significantly deviate from the spectra in a clean (violet) or the total area. Consequently,
the spectra are not significantly influenced by these defects.

The d$I$/d$V$-spectra of YPtBi are mostly featureless, with a $V$-like shape near
$E_{\rm F}$ and a minimum at around $- 10$ meV. Most importantly, we obtain a finite LDOS
around $E_{\rm F}$, which is a clear indication for a considerable amount of surface
states closing the bulk gap.

This experimental result is to be contrasted with the bulk DOS as calculated by DFT
[black dashed line in Fig.\ \ref{Fig3}(b) and Appendix Fig.\ \ref{bulkband}(b)], which
predicts a gap-like behavior near $E_{\rm F}$. The calculations also find a mostly
featureless spectrum and, for negative bias away from $E_{\rm F}$, qualitatively agree with
our d$I$/d$V$-data. However, a proper analysis, and specifically insight into the details
near $E_{\rm F}$, requires a slab calculation, the result of which is put forward in Fig.\
\ref{Fig3}(c). The calculations were conducted with the Green function technique for
semi-infinite systems assuming a (120) surface representing our experiments in a more
realistic way. As discussed in the Appendix \ref{appendD3}, with the modified Becke-Johnson
(mBJ) pseudopotential \cite{becke.johnson.06,tran.blaha.09,camargo.baquero.12}, it is
neither possible to perform slab calculations using reconstructed surfaces nor to obtain
the LDOS properly. Nonetheless, we are able to obtain important pieces of information to
understand our STS results. As shown in Fig.\ \ref{Fig3}(c), a mixture of surface states
(non-trivial and trivial ones) contribute significantly to the spectral weight within
the bulk gap, which is consistent with previous angle-resolved photoemission spectroscopy
results \cite{liu2016observation,hosen2020observation}. In other words, surface states
dominate the LDOS near $E_{\rm F}$, which clearly indicates that the increase of the
LDOS in YPtBi, compared to the bulk DOS, stems directly from those surface states. As
shown in Fig.\ \ref{Fig3}(c), near the $\Gamma$ point at $\sim$300~meV we obtain a high
surface spectral weight of dangling bonds. Those trivial surface states are very similar
to the van Hove singularity at approximately $- 100$~meV that is found for LuPtBi
\cite{chadov2010tunable,yang2017prediction,xiao2021first}.

The results of our d$I$/d$V$-measurements for YPtBi become even more intriguing when
compared to those of YPdBi, Fig.\ \ref{Fig3}(e). There are two striking distinctions
in the d$I$/d$V$-data of YPdBi: (i) Qualitatively, the LDOS exhibits more features,
with a prominent peak at approximately 115~meV. (ii) Importantly, there is a clearly
observable gap of width $\Delta \sim 100$~meV around $E_{\rm F}$. We should note,
however, that the d$I$/d$V$-data for YPdBi do not strictly go all the way to zero, but
remain finite at a very small value, possibly caused by thermal effects.

A comparison of the experimental data with the calculated bulk DOS is only partially
possible, Fig.\ \ref{Fig3}(e). On the one hand, the band gap $\Delta^{theor} \sim
0.15$~eV [see Fig.\ \ref{bulkband}(a) in the Appendix] in the projection to the (120)
plane is comparable to the experimental value. This gap once again confirms the trivial
nature of YPdBi, in which the conduction and valence bands are not inverted. Moreover,
there appear to be no surface states near $E_{\rm F}$ in this case. Most of the
spectral weight coming from surface states is located above $E_{\rm F}$, which is
consistent with our d$I$/d$V$ being almost featureless at negative bias.

\subsection{Comparison between YPtBi and YPdBi}
\label{sec:comp}
The differences observed in the d$I$/d$V$-spectra of YPtBi and YPdBi are intriguing
given the facts that identical surface terminations were investigated (thereby
ruling out the surface reconstruction as the main cause of the differences) and
both compound have very small bulk contributions to the DOS near $E_{\rm F}$. The
slab calculated electronic band structures for the YBi-terminated (120) surface plane,
Figs.\ \ref{Fig3}(c) and (f), suggest a considerable admixture of non-trivial
surface states to the DOS near $E_{\rm F}$ in case of the YPtBi surface, which is
absent for YPdBi. In such a case it is a likely scenario that the topological
surface states are the key component for the differences in the d$I$/d$V$-spectra
near $E_{\rm F}$. The minimum observed close to $V_b \sim -10$ meV in case of YPtBi
may then be linked to the Dirac point.

It is also interesting to note that the peak-like feature at $V_{b} = + 115$ meV is
only observed for YPdBi. Two different origins could be at play to cause this peak.
In the first scenario, which is suggested by our slab calculations, this feature
coincides with the bottom of the conduction band, as shown in Fig. \ref{Fig3}(f) and
discussed in Appendix \ref{appendD3} and Fig.\ \ref{pbe_slab}. Here, the lack of this
peak for YPtBi could naturally be explained by the band inversion in this compound.
As discussed in the introduction, non-trivial surface states emerge thereupon.

An alternative scenario involves the presence of the surface reconstruction. Here, the
peak would be a direct consequence of the enhancement of trivial surface states. However,
in this scenario one should also expect such a peak for YPtBi, which is not observed
experimentally. We emphasize that such a comparison is only possible since results
obtained on \emph{identical} surface terminations (concerning the type and arrangement
of the surface atoms as well as the orientation of the terminating plane) are compared.
The clear difference between the LDOS of both systems near $E_{\rm F}$ favours the
increased spin--orbit coupling (upon going from Pd to Pt) as the source of the
appearance of surface states with topological character. The predicted, strong
modification of these surface states has, to the best of our knowledge, not been
demonstrated by STM/S before and suggests a systematic tunability of topology in
half-Heusler systems. Likely, by choosing the proper surface plane, these properties
also may be accessed through macroscopic (e.g.\ transport) measurements.

Finally, our results, even though obtained at $T = 4.6$~K, may also shed some light
on superconductivity in half-Heusler systems. A well defined gap was found for the
trivial insulator YPdBi. This compound has been reported to have one of the highest
superconducting transition temperatures ($T_{\rm c} \sim 1.6$~K) among the $R$PdBi
family \cite{nakajima2015topological}. If the superconductivity is intrinsic, it would
be interesting to understand how a gapped system can develop a superconducting phase.
In this respect it is interesting to note that, at least for YPtBi, some reports
discuss the possibility of superconductivity being a bulk or a surface property
\cite{butch2011superconductivity,kim2022molecular}. Yet, a finite LDOS, possibly with
Dirac point(s), for YPtBi is not inconsistent with a topological superconductivity
scenario \cite{brydon2016pairing,butch2011superconductivity,gofryk2011magnetic,
pavlosiuk2016superconductivity,kim2021campbell,kim2018beyond,kim2022molecular,
zhou-accept}. It would be interesting to conduct further experiments at mK temperatures
to investigate the origin of superconductivity and its nature in half-Heusler systems.

\section{Conclusion}
In summary, we performed scanning tunneling microscopy/spectroscopy on the
half-Heusler systems YPtBi and YPdBi. By {\it in situ} cleaving the single
crystals at low temperatures we were able to investigate atomically flat areas. Both
materials very likely expose (120) YBi-terminated surfaces with ($2\, \times\, 1$)
reconstructions which induce additional surface states and hence, may complicate
surface spectroscopy. Using STM, we can compare \emph{identical} surface terminations,
thereby ruling out the reconstructions as the main cause for differences in the
spectroscopic results between the two materials. However, we do observe a clear
difference in the LDOS of these compounds: While YPdBi exhibits a gap of
$\sim$100~meV around $E_{\rm F}$, surface states are found for YPtBi without
indication of a gap. Such distinct behavior was not seen by macroscopic measurements
reported in previous studies. Our result provides evidence for the targeted
realization of unusual surface states. DFT calculations are consistent with such
change in the LDOS.

In a more general way, our result can very likely be linked to a spin--orbit
tuning of topology in half-Heusler systems. More importantly, our results emphasize
the key role of surface states near $E_{\rm F}$ in these systems. Exploring planes
such as the (120) surface termination [or even the (001) plane in thin films] appears
as an extremely promising route to obtain a versatile TI with an insulating bulk and
increases the potential of half-Heusler systems for applications.

\section*{Acknowledgments}
We thank E.\ H.\ da Silva Neto, T.\ J.\ Boyle and M.\ Walker for their help and
discussions in the beginning of this project. This work was supported by FAPESP\
(SP-Brazil) Grants No 2020/12283-0, 2018/11364-7, 2017/10581-1, CNPq grant No
311783/2021-0 and CAPES. Work was also supported by the National Science Centre
(NCN, Poland) under Projects No.\ 2021/43/B/ST3/02166 (A.P.).  A.P.\ appreciates
funding within the frame of scholarships of the Minister of Science and Higher
Education of Poland for outstanding young scientists (2019 edition,
No.~818/STYP/14/2019). Work at Los Alamos was supported by the Los Alamos
Laboratory Directed Research and Development program through project 20210064DR.

\appendix
\section{(111) plane in YP\lowercase{t}B\lowercase{i}}
\label{appendA}
\subsection{STM/STS results}
\label{appendA1}
\begin{figure*}[t]
\includegraphics[width=0.82\linewidth]{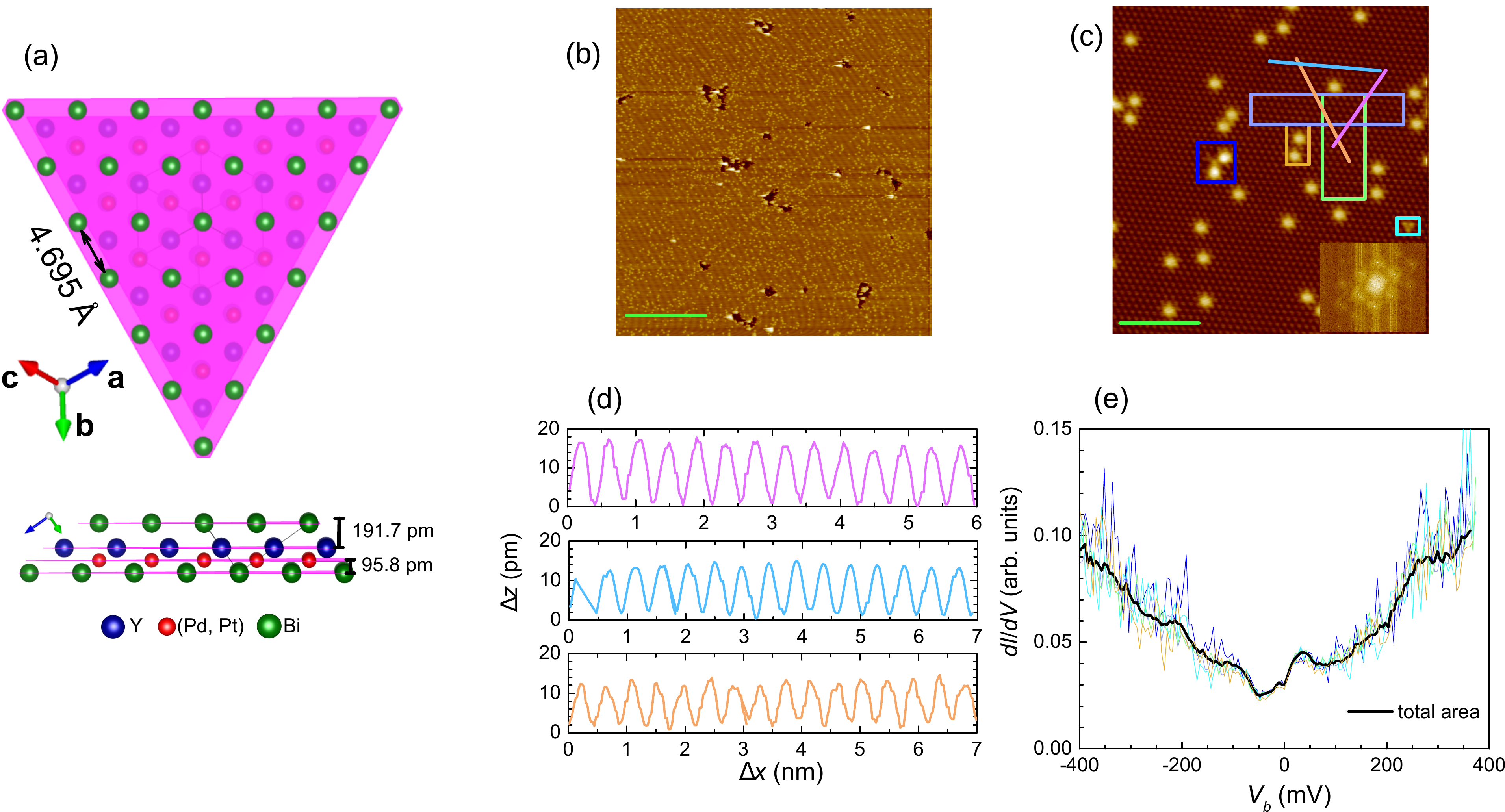}
\unitlength1cm \begin{picture}(0.44,2)
\put(-9.28,7.55){\sffamily {(b)}}
\put(-4.4,7.55){\sffamily {(c)}}
\end{picture}
\caption{(a) The triangular lattice of the (111) plane of the half-Heusler systems and
the height difference between distinct termination planes. (b) $200 \times 200$~nm$^{2}$
STM topography of the (111) plane of YPtBi ($V_{b} = - 300$~mV, $I_{sp} = 0.6$~nA, scale
bar of $50$~nm). (c) Atomically resolved $20 \times 20$~nm$^{2}$ topography on YPtBi
along the (111) surface (scale bar $5$~nm). The lower right inset shows the Fourier
transform. The magenta, blue and orange solid lines represent the direction of the
height scans presented in (d). (e) d$I$/d$V$-spectra averaged over the rectangular areas
of corresponding colors as well as over the total field of view in (c) [black line
in (e), obtained on a $35 \times 35$ grid].} \label{S111}
\end{figure*}
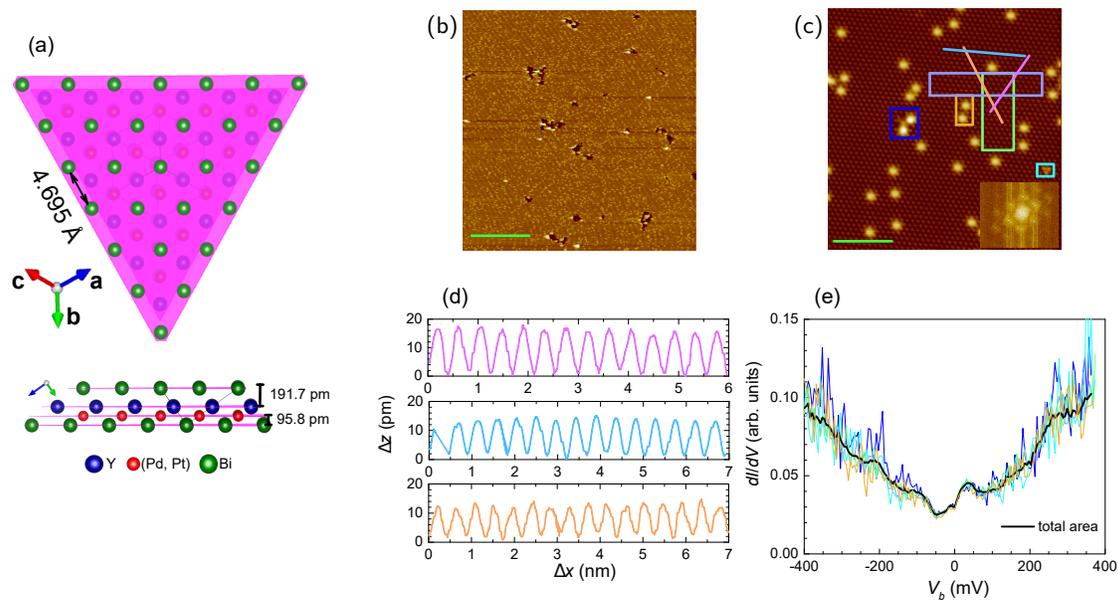

In YPtBi crystals with pyramid-like shape we are also able to cleave along the (111)
plane; the results are summarized in Fig.~\ref{S111}. In this plane, the height
difference between Y and Bi layers is twice as large as the Y-Pt or Pt-Bi layer
spacing, see Fig.~\ref{S111}(a). Furthermore, fewer chemical bonds need to be broken
between Y and Bi layers compared to cleaves involving Pt layers, which should result
in cleaves exposing mostly Y or Bi layers~\cite{liu2016observation}. We found more
easily atomically flat surfaces along the (111) plane when compared to the cleaves
along the (120) plane for YPtBi. Yet, atomically flat surfaces needed to be
extensively searched for, which is not surprising in a cubic system. Importantly,
however, despite great efforts atomically flat areas on a (111) plane could
\emph{not} be found on YPdBi and therefore, the comparison of the LDOS for both
compounds was focused on the (120) plane.

The cleave along the (111) plane should expose either Y- or Bi-terminated triangular
lattices, as shown in Fig.\ \ref{S111}(a). Fig.\ \ref{S111}(b) exhibits a $200 \times
200$~nm$^{2}$ topography. Albeit we were able to locate such large atomically flat areas,
there was quite an amount of adatoms on top of such surfaces. The topography in Fig.\
\ref{S111}(c) zooms into an area of $20 \times 20$~nm$^{2}$. In this case it is possible
to observe in more detail the triangular lattice, which is confirmed by the Fourier
transform presented in the lower right inset. Again, we obtain a moderate amount of
adatoms, which are expected on an unreconstructed surface due to its polar nature. The
establishment of an unreconstructed surface is further corroborated by the distance
\begin{figure}[!ht]
\includegraphics[width=0.95\linewidth]{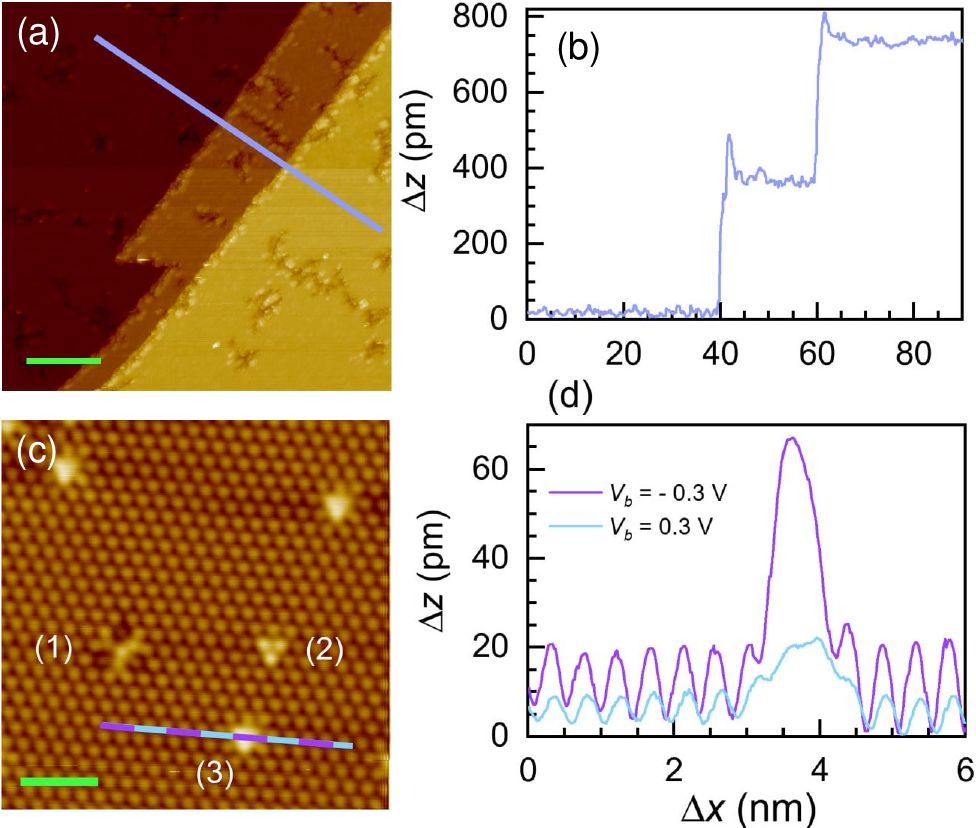}
\caption{(a) $100 \times 100$~nm$^{2}$ STM topography along the (111) plane of YPtBi
($V_{b} = 200$~mV, $I_{sp} = 0.3$~nA, scale bar of $20$~nm). The purple line represents
the line along which the height scan in (b) was obtained. (c) $10 \times 10$~nm$^{2}$
field of view ($V_{b} = - 300$~mV, $I_{sp} = 0.3$~nA, scale bar of $2$~nm). Here, the
three most dominant defects can be recognized, which likely are a triangular vacancy
[defect (1)], a triangle which could be related to a substitution [defect (2)], and 
adatoms [defect (3)]. The latter is supported by the violet and blue height scans in 
(d) obtained for opposite bias voltages along the lines indicated in (c).}
\label{topo111}
\end{figure}
between corrugations, as highlighted in the height scans of Fig.\ \ref{S111}(d). We
obtain a distance between corrugations of $d_{exp}^{(111)} = 0.43(2)$~nm, $0.46(2)$~nm,
and $0.44(2)$~nm for the magenta, blue and orange lines, respectively, which is in
excellent agreement with the theoretical distance between Bi/Y atoms of $d_{theor}^{(111)
\text{Y/Bi}} = 0.47$~nm along the (111) plane. As we will discuss in more detail below,
the exposed surface is likely an unreconstructed Bi-terminated surface.

In Fig.~\ref{S111}(e) we present d$I$/d$V$-spectra, which were obtained within in
\begin{figure*}[t]
\includegraphics[width=0.96\linewidth]{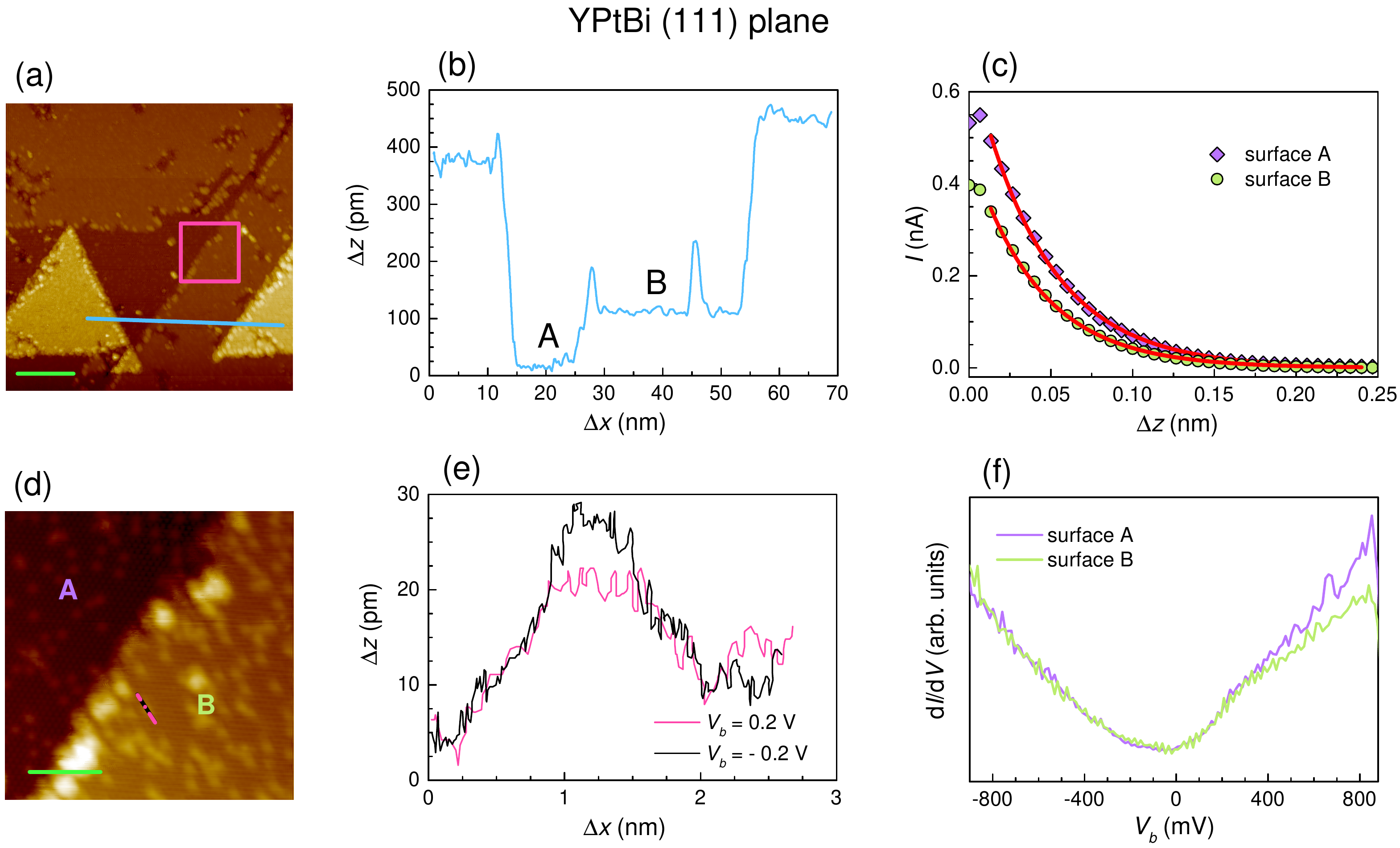}
\caption{(a) $100 \times 100$~nm$^{2}$ STM topography along the (111) plane for YPtBi
($V_{b} = - 200$~mV, $I_{sp} = 0.3$~nA, scale bar of $20$~nm). (b) Height scan along the
blue line shown in (a). The $A$ and $B$ labels denote the surfaces $A$ (Bi-terminated)
and $B$ (likely Pt-terminated). (c) Tunneling current $I$ as a function of the tip-sample
distance $\Delta z$ for the two different surfaces A and B. The red solid lines are
exponential fits as described in the text. (d) Zoom of $20 \times 20$~nm$^{2}$ into the
magenta box shown in (a) (scale bar of $5$~nm). (e) Height scan along the magenta/black
line shown in (d) and for different $V_b$. The lines cross an adatom located in
surface $B$. (f) Averaged d$I$/d$V$-spectra of both surfaces A and B. The averages
were taken over areas of $4 \times 4$~nm$^{2}$ at equally spaced positions on
$50 \times 50$~grids.} \label{step111}
\end{figure*}
the total field of view of Fig.\ \ref{S111}(c) (black line) as well as within the areas
marked by colored rectangles (with the colors corresponding to those of the spectra).
Earlier theoretical calculations indicated a Dirac point buried in the bulk DOS for
either Y- or Bi-terminated surfaces, which was confirmed by angle resolved photoemission
spectroscopy \cite{liu2016observation,hosen2020observation}. Nonetheless, trivial
Rashba-like surface states can be expected due to the presence of dangling bonds
\cite{liu2016observation,kawasaki2018simple}. The observed d$I$/d$V$-spectra are almost
featureless, with a finite DOS at the Fermi level $E_{\rm F}$. Interestingly, the LDOS
obtained at two adatoms [yellow area in Fig.\ \ref{S111}(c)], or at defects [blue area
in Fig.\ \ref{S111}(c)] does not change significantly when compared to the LDOS obtained
on clean surfaces (green and purple areas and spectra) or even to the spectrum averaged
over the total field of view. This is an indication that the surface states are not
affected locally by small amounts of disorder.

In order to provide further evidence to the (111) assignment of the terminating plane
observed in Fig.\ \ref{S111}, we experimentally explored the presence of step edges and
adatoms. Fig.~\ref{topo111}(a) shows a typical 100 x 100 nm$^{2}$ topography along the (111)
plane with two step edges. As can be seen in Fig.\ \ref{topo111}(b), the height difference
between each exposed surface is $h^{(111)}_{exp} \approx 370$~pm. Such a distance is
consistent to either Y--Y, Bi--Bi or Pd--Pd/Pt--Pt surface terminations for which
$h^{(111)}_{theor} = 383$~pm is expected. It is worth to note that there is an
accumulation of adatoms along the edges of the exposed surfaces (small peaks in the
height scan), which reinforces our assumption above that such adatoms play a role in
neutralization of the exposed polar surfaces.

Such an assumption is also corroborated by different apparent heights of the adatom
if measured with different bias $V_b$. Fig.\ \ref{topo111}(c) shows a $10 \times
10$~nm$^{2}$ topography taken with $V_{b} = - 300$~mV. We can observe the three most
numerous defects obtained on these surfaces: a triangular one likely related to a vacancy
[defect (1)], a small triangle which could be related to a substitution in a sub-layer
underneath the exposed surface [defect (2)] and the already mentioned adatoms [defect
(3)]. Fig.\ \ref{topo111}(d) provides the height scan across the adatom position in Fig.\
\ref{topo111}(c) with different values of the applied $V_{b}$ obtained in dual-bias mode
(i.e.\ at exactly the same position). We systematically observe higher heights at the
adatom sites for negative $V_{b}$. For negative (positive) $V_{b}$, the tip will have a
positive (negative) potential with respect to the sample. In this scenario, the tip gets
further away from (closer to) the adatom if it has a more positive charge compared to
the surrounding bulk. We expect a valence of $3+$ for Y, while the other constituents
in YPtBi should have a more negative valence. Consequently, an adatom is much more
likely more positive as its surrounding on a Bi or Pt terminated surface. Since a cleave
exposing Pt is unlikely from the chemical bonding situation discussed above, we speculate
that we obtained a Bi terminated surface in Fig.\ \ref{topo111}.

\subsection{Coexistence of Pt and Bi-terminated surfaces}
\label{appendA2}
As already pointed out, the majority of the YPtBi samples which successfully cleaved
along the (111) plane exposed the same termination within the field of view (even when
step edges were included, cf.\ Fig.\ \ref{topo111}). However, in one particular field of
view we were able to observe a coexistence of two differently terminated surfaces. Fig.\
\ref{step111}(a) shows this $100 \times 100$~nm$^{2}$ topography. In the height scan
explored in Fig.\ \ref{step111}(b) we obtain a step height between two consecutive
surfaces (labeled $A$ and $B$) of $\approx 95$~pm. This value is much smaller than the
expected step height $h^{(111)}_{theor}$ = 383 pm for identical terminations. Indeed,
it is close to the Pt--Bi (or Pt--Y) layer distance, which is $95.8$ pm [Fig.\
\ref{S111}(a)]. Such a step height necessarily implies that one of the surfaces
\emph{has} to have a Pt termination.

The difference between those two surface terminations is also manifested by two
different heights of the tunneling barrier $\Phi$, which is closely related to the work
function $\Phi_{s}$ of the sample (note that also the tip work function $\Phi_{t}$
enters into $\Phi$). $\Phi$ can be obtained from an analysis of $I$ as a function of the
tip-sample distance $\Delta z$. For clean surfaces, $I(\Delta z) \propto \exp (-2 \kappa
\, \Delta z)$ with $\kappa^{2} = 2m_{e} \Phi /\hbar^{2}$, where $m_{e}$ is the bare
electron mass and $V_{b} \ll \Phi_{s,t}$. Fig.\ \ref{step111}(c) represents
$I(\Delta z)$ curves for surfaces $A$ and $B$, which are identified in Figs.\
\ref{step111}(b) and (d). By fitting the $I(\Delta z)$ curves, red lines in Fig.\
\ref{step111}(c), we obtain $\Phi_{A} \approx 4.9$~eV and $\Phi_{B} \approx 5.5$~eV. A
fair comparison here is to look at the values of the elemental materials. For Y, Bi and
Pt, $\Phi_{s} = 3.1$~eV, $4.22$~eV, and $5.65$~eV, respectively. Comparing to our
obtained results, one may speculate that the highest obtained value, i.e.\ $\Phi_{B}$,
is unlikely from an Y-terminated surface and, conversely, the lower value $\Phi_{A}$
does not stem from a Pt-terminated surface. As one of the two surfaces ($A$ or $B$)
has to be Pt-terminated, it is likely surface $B$.
\begin{figure}[!t]
\includegraphics[width=\columnwidth]{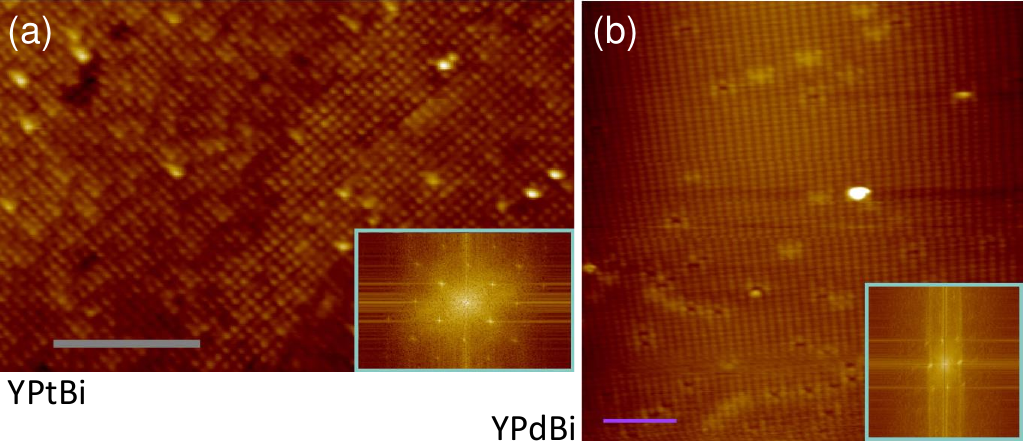}
\caption{(a) 25 x 39 nm$^{2}$ and (b) 30 x 30 nm$^{2}$ topographies along the (120) plane
for YPtBi ($V_{b} = -300$ mV, $I_{sp}$ = 0.7 nA, scale bar of 10 nm) and YPdBi ($V_{b} =
-200$ mV, $I_{sp}$ = 0.6 nA, scale bar of 5 nm), respectively. The right insets show the
Fourier transform of the respective topography.} \label{FT120}
\end{figure}

A closer look at height scans across defects can also be informative with respect to the
assignment of those two distinct surfaces. A zoom into the magenta box of Fig.\
\ref{step111}(a) is given in Fig.\ \ref{step111}(d). According to Fig.\ \ref{step111}(b),
we identify the differently terminated surfaces as $A$ and $B$ in the $20 \times 20$
nm$^{2}$ topography. The height scans across a defect at surface $B$ for opposite $V_{b}$
signs are shown in Fig.\ \ref{step111}(e). They were taken at the positions highlighted by
the magenta and black lines in Fig.\ \ref{step111}(d). It is straightforward to note that
the difference of the defect height for opposite $V_{b}$-values is much smaller on this
surface when compared to the adatom height at surface $A$ (which was assigned as
Bi-terminated), cf.\ Fig.\ \ref{topo111}(d). In consequence, the defect investigated in
Fig.\ \ref{step111}(e) is very likely located in a sub-surface layer, i.e., in a layer
underneath the exposed surface.
\begin{figure}[b]
\includegraphics[width=0.68\linewidth]{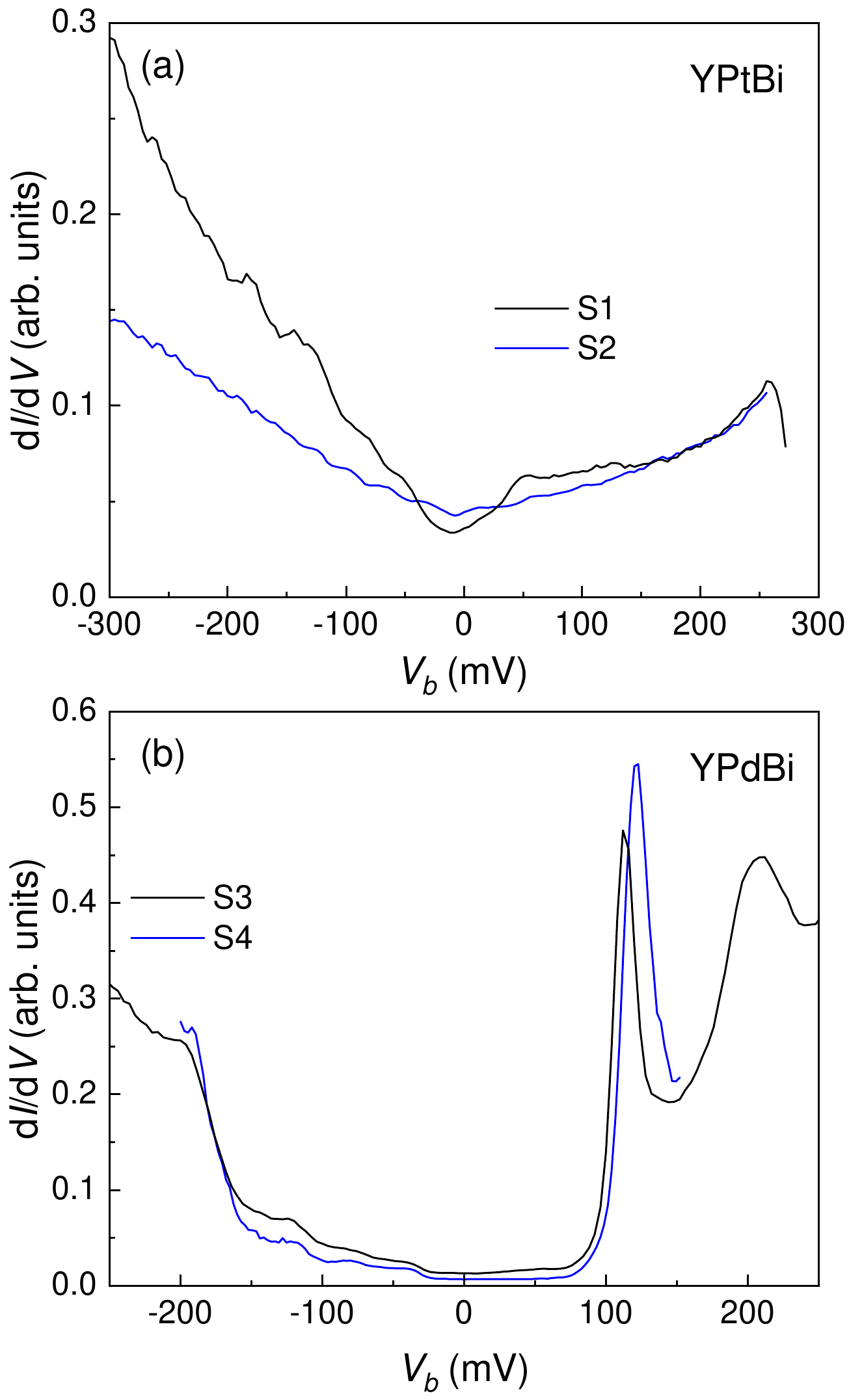}
\caption{Comparison of results obtained on different samples of (a) YPtBi and (b)
YPdBi. Clearly, the d$I$/d$V$-curves of a given material are well reproduced. All
spectra are averages over certain areas; for S1 and S3 they are described in Fig.\
\ref{Fig3}. The averages were taken within a $20 \times 10$ nm$^{2}$ field of view
($V_{b} = -$300~mV, $I_{sp}$ = 0.7 nA) for spectra S2, and a $4 \times 4$ nm$^{2}$ field
of view ($V_{b} = -200$~mV, $I_{sp}$ = 0.6 nA) for S4.} \label{STScomp}
\end{figure}

Finally, Fig.\ \ref{step111}(f) represents d$I$/d$V$-spectra as a function of $V_{b}$ for
both surfaces. Surprisingly, there is only a small difference at higher positive
$V_{b}$-values between the spectra of the two differently terminated surfaces, suggesting
that the LDOS is dominated by bulk and trivial surfaces state contributions in both cases.

\section{Fourier transform of the $(120)$ planes}
\label{appendB}
A more accurate extraction of the distance between corrugations can be achieved by
analyzing the Fourier transform of larger areas. In the case of YPtBi, one large flat
area that we were able to obtain is shown in Fig.~\ref{FT120}(a). From the Fourier
transform (inset), we obtained $d_{exp}^{(120)Pt}$ = 0.67(3) and 0.70(3) nm. For
YPdBi, a large, atomically flat area is presented in Fig.\ \ref{FT120}(b). Here, the
Fourier transform yielded $d_{exp}^{(120)Pd}$ = 0.61(3) and 0.72(3) nm. It is worth
to notice that the Fourier transforms even have a slightly rectangular shape [see
also Fig.\ \ref{Fig2}(i)], consistent with the asymmetry between the distance of
corrugations.
\begin{figure}[b]
\includegraphics[width=0.89\columnwidth]{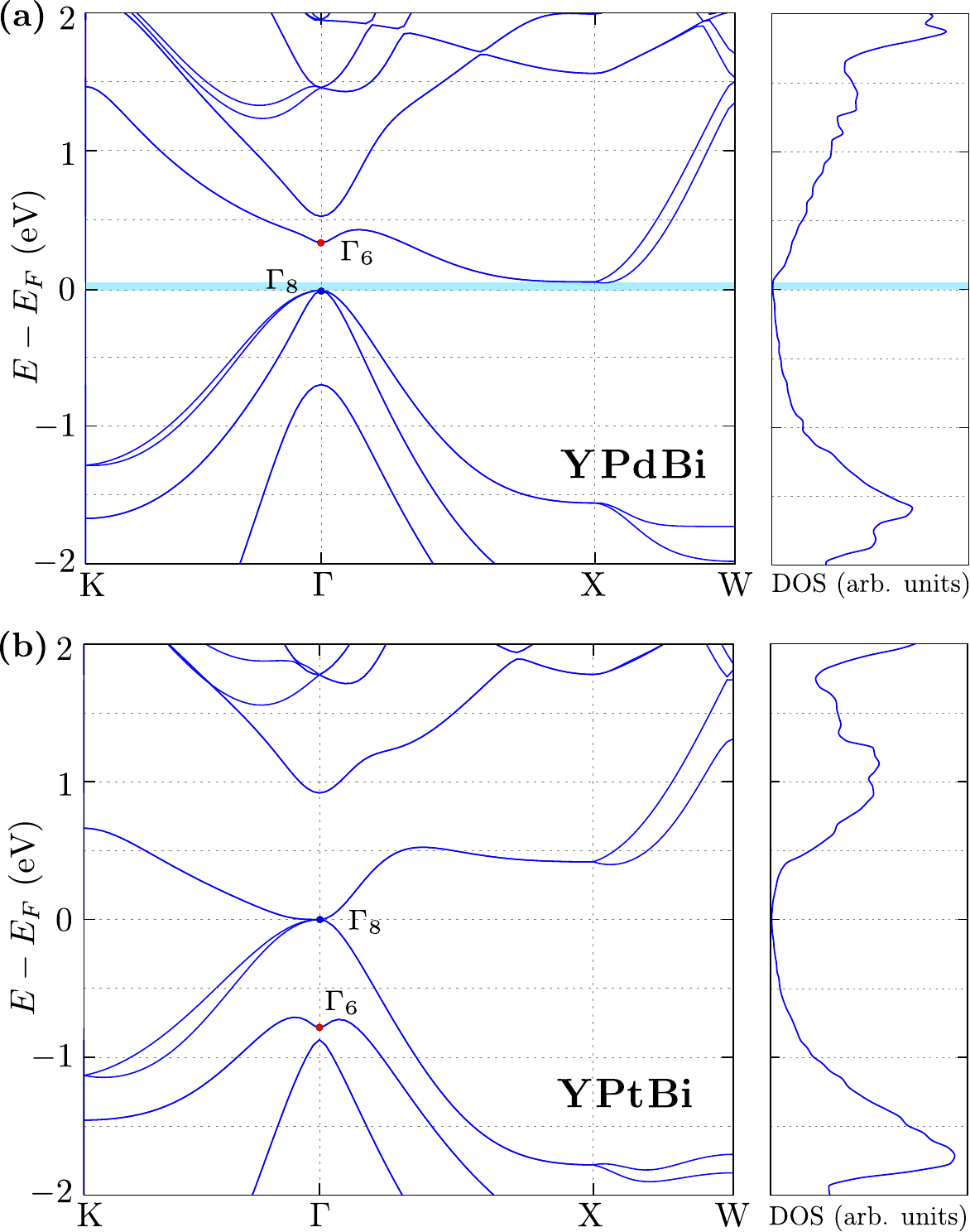}
\caption{Electronic band structures and density of states (DOS) for bulk (a) YPdBi
and (b) YPtBi. The $\Gamma_{6}$ and $\Gamma_{8}$ bands are marked in the figure. The
light-blue stripe in (a) visualizes the gap of $\sim$0.15 eV width.}
\label{bulkband}
\end{figure}

\section{Reproducibility of the spectra}
\label{appendC}
As mentioned in Sec.\ \ref{sec:Methods}, four YPdBi and three
YPtBi samples were cleaved \textit{in situ} and subsequently investigated by STM/STS.
In order to provide support for the reproducibility of our data, specifically the
spectra, we here exemplify results obtained on different samples of YPtBi as well as
YPdBi. For comparison, we also reproduced in Fig.\ \ref{STScomp} the spectra
of Fig.\ \ref{Fig3}, marked S1 and S3, respectively. Clearly, the spectra of different
samples of a given material compare well, and all the main features are reproduced.
Most important for the main conclusion of our investigation, the reduction of the LDOS
at $E_{\rm F}$ is clearly reproduced upon going from YPtBi to YPdBi.
\begin{figure}[t]
\includegraphics[width=0.72\linewidth]{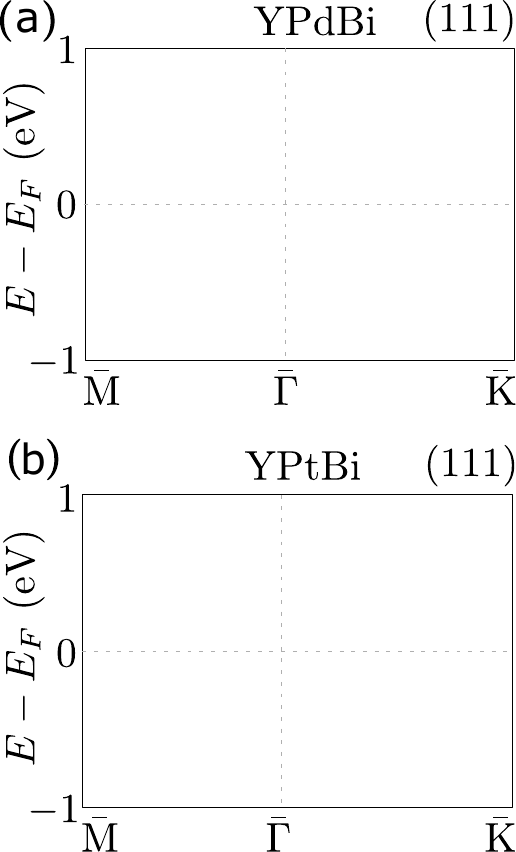}
\caption{Surface spectral functions of (a) YPdBi and (b) YPtBi for the (111) surface
planes.} \label{Green}
\end{figure}

\section{Details of Bulk and surface band structure calculations}
\label{appendD}
\subsection{Bulk band structure calculations}
We were able to reproduce the bands at $\Gamma$ point using the modified Becke-Johnson
(mBJ) pseudopotential, see Fig.~\ref{bulkband}, as given in~\cite{al2010topological,
feng.xiao.10}. Here, the topological properties can be described by $\Delta E =
E_{\Gamma_{6}} - E_{\Gamma_{8}}$, which is negative for systems with band inversion
\cite{feng.xiao.10} (as mentioned in the introduction). Indeed, for YPdBi we find
$\Delta E \simeq +0.35$~eV, while for YPtBi we obtain approximately
$-0.79$~eV. In other words, our calculations specify YPtBi as a zero-gap
semiconductor. Both values are similar to those reported earlier
\cite{feng.xiao.10,chadov2010tunable,al2010topological}. It is worth to point out
that even though different pseudopotentials might result in similar densities of
states, some of those pseudopotentials do not reproduce correctly the band
inversion \cite{meineert.16}. Finally, these results depend strongly on the lattice
constant \cite{chadov2010tunable}.

\subsection{Slab calculations for the (111) plane}
\label{appendD2}
The surface spectral functions, calculated within the Green function technique for
semi-infinite systems, for the (111) surface planes of YPdBi and YPtBi are presented
\begin{figure}[t]
\includegraphics[width=0.75\linewidth]{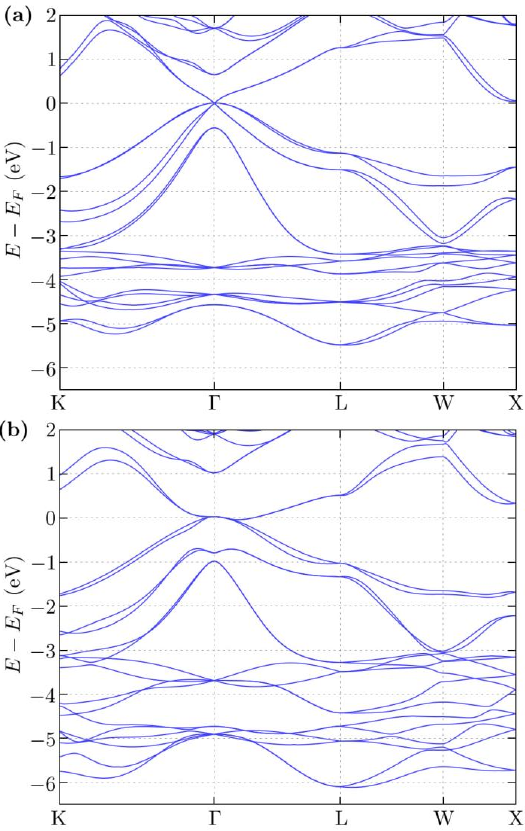}
\caption{The electronic band structure of (a) bulk YPdBi and (b) YPtBi obtained from
DFT GGA PBE calculations.} \label{pbe_bands}
\end{figure}
in Figs.\ \ref{Green}(a) and (b), respectively [cf.\ Figs.\ \ref{Fig3}(c) and (f) for
the (120) plane]. In the case of a Bi-terminated (111) surface of YPdBi, there is no
Dirac point at the $\bar{\Gamma}$ point, as expected [Fig.\ \ref{Green}(a)]. In the
YPtBi case, the Dirac point appears below the Fermi level $E_{\rm F}$, as seen
in Fig.\ \ref{Green}(b). This result is consistent with previous DFT calculations and
ARPES experiments \cite{liu2016observation,hosen2020observation}. It is worth to note
that our slab calculations improve our understanding of the lack of changes in the LDOS
near defects in this plane, cf.\ Fig.~\ref{S111}(e). As discussed in Sec.\
\ref{sec:spec}, the bulk LDOS has some impact in our data. Near $E_{\rm F}$, we
find trivial and topological bands, which may complicate the scattering process even
further. Therefore, the change in the LDOS due to trivial surface states may be hard
to detect within this energy window.\\

\subsection{Direct slab band structure calculations}
\label{appendD3}
Due to the technical limitations of the mBJ potential implemented in {\sc vasp}, it
cannot be used directly to study the slab band structure. As such, to present the main
impact of the surface reconstruction on the electronic band structure, we performed DFT
calculations using GGA with Perdew--Burke--Enzerhof (PBE) parametrization
\cite{perdew.burke.96}. Here, we should emphasize that this approach (DFT with GGA
+ PBE) does not correctly reproduce the bulk band structure of half-Heusler compounds,
a problem well reported in the literature \cite{becke.johnson.06,tran.blaha.09,
camargo.baquero.12}. This is reflected in the absence of the band gap for YPdBi
[cf.\ Fig.\ \ref{bulkband}(a) and Fig.\ \ref{pbe_bands}(a)] or incorrect band
curvatures around the $\Gamma$ point in the band structure of YPtBi [cf. Fig.\
\ref{bulkband}(b) and Fig.\ \ref{pbe_bands}(b)]. Nevertheless, this type of calculation
can be used to present the main features of the band structure of (120) surfaces with
reconstruction.
\begin{figure*}[t]
\hspace*{1.0cm}without reconstruction \hspace*{5.2cm}with reconstruction
\includegraphics[width=0.96\linewidth]{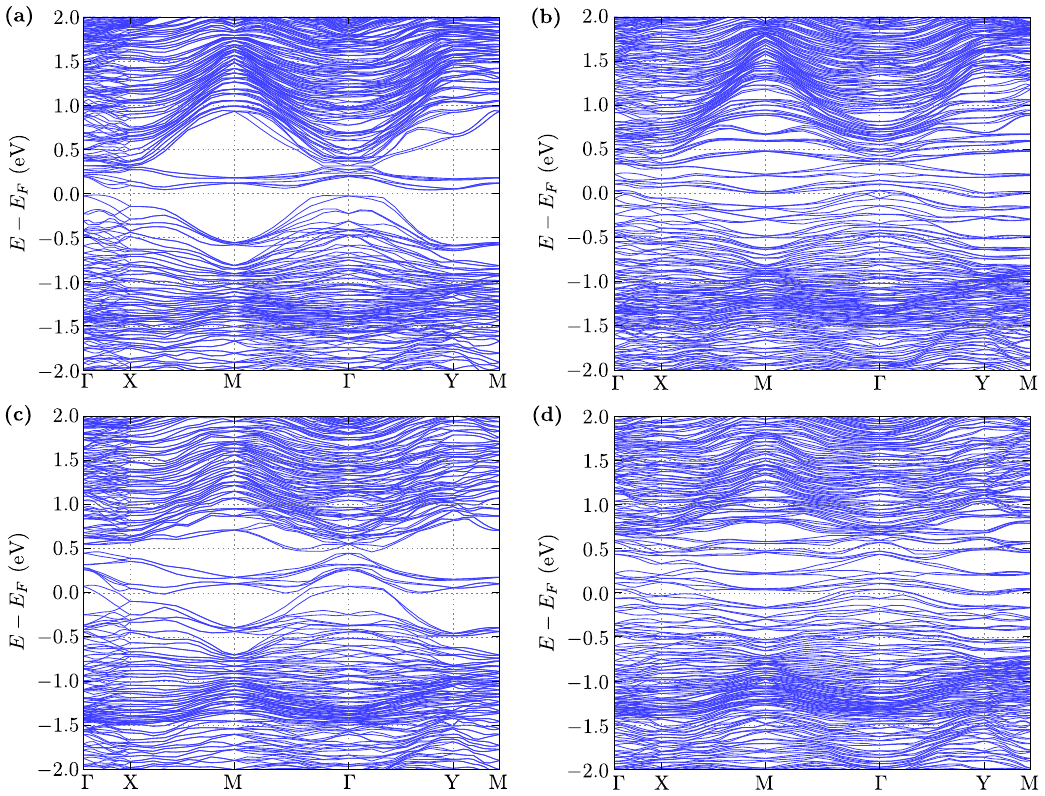}
\begin{turn}{90}\hspace*{2.4cm}YPtBi \hspace*{5.4cm}YPdBi\end{turn}
\caption{The electronic band structure of slab YPdBi (top panels) and YPtBi (bottom
panels) obtained from DFT GGA PBE calculations. Left and right panels correspond to the
(120) surface without and with surface reconstruction, respectively.} \label{pbe_slab}
\end{figure*}

For the simulation of the (120) surface band structure, we constructed slab models
containing 3 layers of the discussed compounds (mostly 28 formula units). The
reconstructions of the surface was introduced ``by hand'', i.e., by removing some
atoms from the surfaces. From the self-consistently found charge distributions, the
STM simulations were computed [see Figs.\ \ref{Fig2}(j) and (k)]. Similarly, the
electronic band structure for both compounds are presented in Fig.~\ref{pbe_slab}
where the results obtained for surfaces without and with surface reconstruction are
presented on the left and right panels, respectively. For both YPdBi and YPtBi (120)
surfaces without reconstruction (left panels in Fig.~\ref{pbe_slab}), we observed the
realization of the surface states above $E_{\rm F}$. The introduction of surface
reconstructions on the (120) surface led to a multiplication of the surface states,
an effect that is well visible at the M points in the right panels of Fig.\
\ref{pbe_slab}. The additional, ``extra'' surface states come from the hanging
(non-bonded) orbitals in the surface plane, due to the absence of some atoms on the
surface (i.e.\ the surface reconstruction). This main feature of the band structure
for the reconstructed surface, i.e.\ the surface states multiplication, is expected
to also be present in case of a ``correctly'' obtained band structure, i.e.\ if
calculated with mBJ potential.

\end{document}